\UseRawInputEncoding 
\documentclass[twocolumn]{aastex61}

\newcommand{\lsim}{\raisebox{-0.13cm}{~\shortstack{$<$ \\[-0.07cm] $\sim$}}~}
\newcommand{\gsim}{\raisebox{-0.13cm}{~\shortstack{$>$ \\[-0.07cm] $\sim$}}~}

\received{...}
\revised{...}
\accepted{...}
\submitjournal{ApJ}

\shorttitle{An ALMA galaxy signposting a MUSE galaxy group at {\lowercase{$z=4.3$}}}
\shortauthors{Caputi et al.}

\begin{document}

\title{ALMA Lensing Cluster Survey: an ALMA galaxy signposting a MUSE galaxy group at {\lowercase{$z=4.3$}} behind ``El Gordo''}

\correspondingauthor{K. I. Caputi}
\email{karina@astro.rug.nl}

\author{K. I. Caputi}
\affil{Kapteyn Astronomical Institute, University of Groningen, P.O. Box 800, 9700AV Groningen, The Netherlands}
\affil{Cosmic Dawn Center (DAWN), Copenhagen, Denmark}

\author{G. B. Caminha}
\affil{Kapteyn Astronomical Institute, University of Groningen, P.O. Box 800, 9700AV Groningen, The Netherlands}

\author{S. Fujimoto}
\affil{Cosmic Dawn Center (DAWN), Copenhagen, Denmark}
\affil{Niels Bohr Institute, University of Copenhagen, Lyngbyvej 2, DK-2100, Copenhagen, Denmark}

\author{K. Kohno}
\affil{Institute of Astronomy, Graduate School of Science, University of Tokyo, 2-21-1 Osawa, Mitaka, Tokyo 181-0015, Japan}
\affil{Research Center for the Early Universe, School of Science, The University of Tokyo, 7-3-1 Hongo, Bunkyo-ku, Tokyo 113-0033, Japan}

\author{F. Sun}
\affil{Steward Observatory, University of Arizona, 933 N. Cherry Ave, Tucson, AZ 85721, USA}

\author{E. Egami}
\affil{Steward Observatory, University of Arizona, 933 N. Cherry Ave, Tucson, AZ 85721, USA}

\author{S. Deshmukh}
\affil{Kapteyn Astronomical Institute, University of Groningen, P.O. Box 800, 9700AV Groningen, The Netherlands}

\author{F. Tang}
\affil{Kapteyn Astronomical Institute, University of Groningen, P.O. Box 800, 9700AV Groningen, The Netherlands}

\author{Y. Ao}
\affil{Purple Mountain Observatory and Key Laboratory for Radio Astronomy, Chinese Academy of Sciences, Nanjing, China}
\affil{School of Astronomy and Space Science, University of Science and Technology of China, Hefei, Anhui, China}

\author{L. Bradley}
\affil{Space Telescope Science Institute, 3700 San Martin Drive, Baltimore, MD 21218, USA}

\author{D. Coe}
\affil{Space Telescope Science Institute, 3700 San Martin Drive, Baltimore, MD 21218, USA}

\author{D. Espada}
\affil{SKA Organisation, Lower Withington, Macclesfield, Cheshire SK11 9DL, UK}

\author{C. Grillo}
\affil{Dipartimento di Fisica, Universit\`a degli Studi di Milano, via Celoria 16, I-20133 Milano, Italy}
\affil{Dark Cosmology Centre, Niels Bohr Institute, University of Copenhagen, Lyngbyvej 2, DK-2100 Copenhagen, Denmark}
\affil{INAF - IASF Milano, via A. Corti 12, I-20133 Milano, Italy}

\author{B. Hatsukade}
\affil{Institute of Astronomy, Graduate School of Science, University of Tokyo,  2-21-1 Osawa, Mitaka, Tokyo 181-0015, Japan}

\author{K. K. Knudsen}
\affil{Department of Space, Earth and Environment, Chalmers University of Technology, Onsala Space Observatory, SE-43992 Onsala, Sweden}

\author{M. M. Lee}
\affil{Max-Planck-Institut für Extraterrestrische Physik (MPE), Giessenbachstr., D-85748 Garching, Germany} 

\author{G. E. Magdis}
\affil{Cosmic Dawn Center (DAWN), Copenhagen, Denmark}
\affil{DTU-Space, Technical University of Denmark, Elektrovej 327, DK-2800 Kgs. Lyngby, Denmark}
\affil{Niels Bohr Institute, University of Copenhagen, Lyngbyvej 2, DK-2100 Copenhagen \O, Denmark}

\author{K. Morokuma-Matsui}
\affil{Institute of Astronomy, Graduate School of Science, University of Tokyo, 2-21-1 Osawa, Mitaka, Tokyo 181-0015, Japan}

\author{P. Oesch}
\affil{Department of Astronomy, University of Geneva, Ch. des Maillettes 51, 1290 Versoix, Switzerland}

\author{M. Ouchi}
\affil{National Astronomical Observatory of Japan, 2-21-1 Osawa, Mitaka, Tokyo 181-8588, Japan}
\affil{Institute for Cosmic Ray Research, The University of Tokyo, 5-1-5 Kashiwanoha, Kashiwa, Chiba 277-8582, Japan}
\affil{Kavli Institute for the Physics and Mathematics of the Universe (Kavli IPMU, WPI), The University of Tokyo, 5-1-5 Kashiwanoha, Kashiwa, Chiba 277-8583, Japan}

\author{P. Rosati}
\affil{Dipartimento di Fisica e Scienze della Terra, Università degli Studi di Ferrara, Via Saragat 1, I-44122 Ferrara, Italy}
\affil{INAF - Osservatorio di Astrofisica e Scienza dello Spazio, via Gobetti 93/3, I-40129 Bologna, Italy}

\author{H. Umehata}
\affil{RIKEN Cluster for Pioneering Research, 2-1 Hirosawa, Wako-shi, Saitama 351-0198, Japan}
\affil{Institute of Astronomy, The University of Tokyo, 2-21-1 Osawa, Mitaka, Tokyo 181-0015, Japan}

\author{F. Valentino}
\affil{Cosmic Dawn Center (DAWN), Copenhagen, Denmark}
\affil{Niels Bohr Institute, University of Copenhagen, Lyngbyvej 2, DK-2100, Copenhagen, Denmark}

\author{E. Vanzella}
\affil{INAF - Osservatorio di Astrofisica e Scienza dello Spazio, via Gobetti 93/3, I-40129 Bologna, Italy}

\author{W.-H. Wang}
\affil{Institute of Astronomy and Astrophysics, Academia Sinica , No. 1, Sec. 4, Roosevelt Rd. 11F of Astro-Math Building, Da-An District, Taipei 10617, Taiwan}

\author{J. F. Wu}
\affil{Space Telescope Science Institute, 3700 San Martin Drive, Baltimore, MD 21218, USA}

\author{A. Zitrin}
\affil{Department of Physics, Ben-Gurion University, Be'er-Sheva 84105, Israel}




\begin{abstract}

We report the discovery of a Multi Unit Spectroscopic Explorer (MUSE) galaxy group at $z=4.32$ lensed by the massive galaxy cluster ACT-CL J0102-4915 (aka {\it El Gordo}) at $z=0.87$, associated with a 1.2~mm source which is at a $2.07\pm0.88 ~\rm \, kpc$ projected distance from one of the group galaxies. Three images of the whole system appear in the image plane. The 1.2~mm source has been detected within the Atacama Large Millimetre/submillimetre Array  (ALMA) Lensing Cluster Survey (ALCS). As this ALMA source is undetected at wavelengths $\lambda < 2 \, \rm \mu m$, its redshift cannot be independently determined, however, the three lensing components indicate that it belongs to the same galaxy group at $z=4.32$. The four members of the MUSE galaxy group have low to intermediate stellar masses ($\sim 10^7-10^{10} \, \rm M_\odot$) and star formation rates (SFRs) of $0.4-24 \, \rm M_\odot/yr$, resulting in high specific SFRs (sSFRs) for two of them, which suggest that these galaxies are growing fast (with stellar-mass doubling times of only $\sim 2\times 10^7$~years).  This high incidence of starburst galaxies is likely a consequence of interactions within the galaxy group, which is compact and has high velocity dispersion. Based on the magnification-corrected sub-/millimetre continuum flux density and estimated stellar mass, we infer that the ALMA source is classified as an ordinary ultra-luminous infrared galaxy (with associated dust-obscured SFR$\sim 200-300 \, \rm M_\odot/yr$) and lies on the star-formation main sequence. This reported case of an ALMA/MUSE group association suggests that some presumably isolated ALMA sources are in fact signposts of richer star-forming environments at high redshifts.

 \end{abstract}




\section{Introduction} \label{sec:intro}

Lensing fields are excellent systems to study high-redshift galaxies to fainter limits than what is typically possible with current telescopes. Several observational campaigns \citep{pos12,lot17,coe19} have targeted lensing clusters with major observatories, such as the \textit{Hubble Space Telescope (HST)}, enabling a large number of studies of the cluster members, as well as their magnified background sources.

The spectroscopic follow up of these background sources has proven especially useful to reveal the properties of faint galaxies up to redshift $z\sim 6-7$. In addition to the study of pre-selected high-$z$ candidates, observations conducted with the Multi Unit Spectroscopic Explorer (MUSE) on the the Very Large Telescope (VLT) have allowed for significant serendipitous discoveries at high redshifts \citep[e.g.,][]{kar15,cam16,kar17,van17a}, including a few galaxy groups at redshifts $z>3$ . These galaxy groups have been found in association with diffuse Lyman-$\alpha$ nebulae or CIV absorption systems \citep[e.g.][]{cam16,van17b,dia20}. Here,  we report the discovery of a group of Lyman-$\alpha$ emitters/absorbers at $z=4.3$ in association with an Atacama Large Millimetre/submillimetre Array (ALMA) sub-millimetre source. 

For a long time, it has been considered that sub-/millimetre sources could trace active sites with multiple star-forming galaxies \citep[e.g.][]{cha09,mar18}.  To date, this idea has been mainly supported by the discovery of galaxy groups and proto-clusters dominated by bright sub-millimetre galaxies \citep[][]{rie14,ume15,mil18,ote18}. Some studies have reported cases of intrinsically fainter sub-millimetre sources whose flux densities is significantly boosted by gravitational lensing and belong to galaxy groups up to $z\sim3$ \citep[e.g.,][]{bor04,kne04,gar05,ber07,mac14}.  The system that we study here constitutes another example of this kind, albeit at a higher redshift $z=4.32$. 

Studying the properties of member galaxies in these mixed groups with sub-/millimetre sources and Lyman-$\alpha$ emitters/absorbers can help understand the consequences of close galaxy interactions. The majority of star-forming galaxies at different redshifts are located in the so-called star-formation main sequence \citep[e.g.,][]{noe07,spe14,bis18} on the star formation rate (SFR) versus stellar mass $M^\ast$ plane. The fraction of starburst galaxies, i.e. galaxies with significantly enhanced star formation activity with respect to their past average SFR, is known to vary with stellar mass and redshift \citep[e.g.][]{rod11,sar12,cap17,bis18}. In any case, they range between a few and $\sim15\%$  of all known galaxies up to $z\sim5$. This relative importance between main sequence and starburst galaxies might be different in galaxy groups, where galaxy interactions could favour starburst episodes, but this is still poorly understood. The galaxy group that we study here allows us to investigate the importance of starbursts within compact galaxy groups at high redshifts.

Throughout this paper we adopt a cosmology  with $\rm H_0=70 \,{\rm km \, s^{-1} Mpc^{-1}}$, $\rm \Omega_M=0.3$ and $\rm \Omega_\Lambda=0.7$. All magnitudes in this paper are total and are expressed in the AB system \citep{oke83}. Stellar masses and SFRs refer to a \citet{cha03} initial mass function.

\section{Datasets} \label{sec:data}

Our study is based on the analysis of the publicly available MUSE data for the massive galaxy cluster ACT-CL J0102-4915 (aka {\it El Gordo}) at $z=0.87$. {\it El Gordo} is a well-studied galaxy cluster, whose total mass is estimated to be a few $\times 10^{15} \, \rm M_\odot$ \citep{men12,zit13,cer18}. The MUSE data on  ACT-CL J0102-4915 was taken under the ESO programme ID 0102.A-0266 (P.I. G. B. Caminha) and have a final depth of
$\approx$2.3 hours on target using the GALACSI adaptive optics system. The data was processed using the reduction pipeline version 2.6 \citep{wei14}, following all standard corrections and calibrations to create the stacked data-cube. The final data covers the wavelength range $\rm 4700\AA - 9350\AA$ and the measured point-spread function full width at half maximum (FWHM) on the white images is $\approx 0.7\arcsec$. The detailed description of the data will be presented in \citet{cam21}.

The ALMA Band-6 (1.1-1.4~mm) data for {\it El Gordo} has been obtained as part of the ALMA Lensing Cluster Survey (ALCS) Large Programme (ID 2018.1.00035.L; P.I. K. Kohno), which has been carried out in ALMA's cycles 6 and 7. The data have been reduced with the scripts provided by the ALMA observatory, using CASA versions 5.4.0 and 5.6.1 for cycles 6 and 7 data, respectively, following the standard pipeline.

The natural-weighted continuum map achieves a sensitivity of 71.9 $\rm \mu Jy$/beam and a synthesized beam size (FWHM) of $1.17'' \times 0.91''$.  The details of the ALCS programme description and the data reduction and imaging are presented in Kohno et al. (in prep.) and Fujimoto et al. (in prep.), respectively. 

We note that the three components of the ALCS lensed source that we discuss here have also been detected in an earlier ALMA programme (P.I. A. Baker) and reported as three different sources (Wu et al. 2018).

Finally, for the analysis of the ALMA source we also made use of \textit{Herschel Space Telescope} Spectral and Photometric Imaging Receiver (SPIRE) data \citep{gri10}, obtained as a part of the \textit{Herschel} Lensing Survey \citep[HLS;][]{ega10} at a relatively shallow depth ($\mathrm{r.m.s.}=11.8$\,mJy/beam at 250\,\micron). The data were reduced with the standard reduction pipeline described in \citet{raw16}.  We extracted the SPIRE flux densities of the ALCS sources using an iterative point-spread-function fitting algorithm based on ALMA positional priors (Sun et al., in prep.).

\section{An ALMA bright galaxy within a lensed MUSE galaxy group at $\lowercase{z}=4.32$} \label{sec:group}

\begin{figure*}[ht!]
\center{
\includegraphics[width=1.0\linewidth, keepaspectratio]{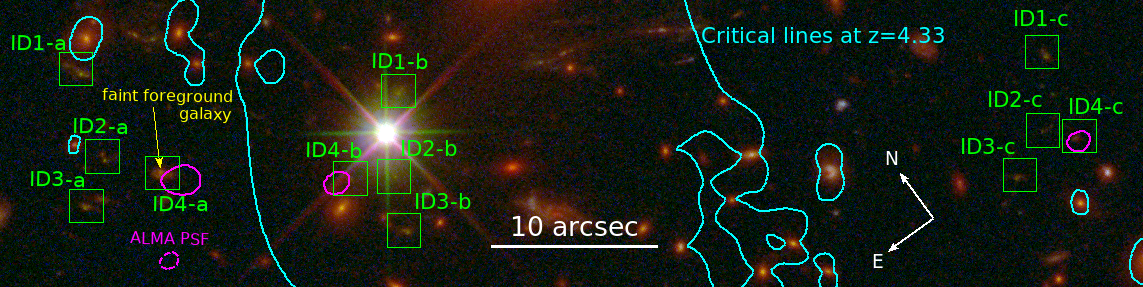}
\caption{\textit{HST} colour image showing the three lensed images of the MUSE/ALMA galaxy group at $z=4.32$ behind the massive galaxy cluster {\it El Gordo} at $z=0.87$ (green squares).  The regions with ALMA continuum emission  (corresponding to the three images of the ALMA source) are delimited with magenta solid lines, while the ALMA beam size is indicated with a magenta dashed-line contour. The critical lines from the strong-lensing best-fit model at $z=4.33$ are also shown (cyan contours). The sources labeled as \#5 and \#6 in Table~\ref{tab:sumtab} are out of the field shown in this Figure. } \label{fig:system}
}
\end{figure*}

\subsection{System description}\label{subsec:sys}

The MUSE data clearly shows the presence of a group of four galaxies (ID \#1 to \#4) at $z_{\rm spec}=4.32-4.33$, which is triply lensed (Fig.~\ref{fig:system}). The lensing model used in this paper, which is based on a large number of spectroscopically confirmed multiple images,  will be presented in a companion paper \citep{cam21}.  As derived from this model, all these group galaxies are found within a physical distance of $30 ~\rm \, kpc$ (in the source plane).  In addition, another galaxy pair (ID \#5 and \#6), with no multiple images, is found behind {\it El Gordo} at $z=4.32$, but is about 300~kpc away of the other, four-member galaxy group (which we call `main group' hereafter). Their connection to the main group is plausible, but not secure, so we will not discuss it further here. The coordinates and properties of all these sources are listed in Table~\ref{tab:sumtab}.
 
\citet{zit13} and \citet{die19} have previously identified a few of these multiply lensed galaxies from \textit{HST} data, but had no spectroscopic redshifts for them. Nevertheless, their reported photometric/lensing-model redshifts were very close to our spectroscopic determinations.

\begin{figure*}[ht!]
\center{
\includegraphics[width=0.8\linewidth, keepaspectratio]{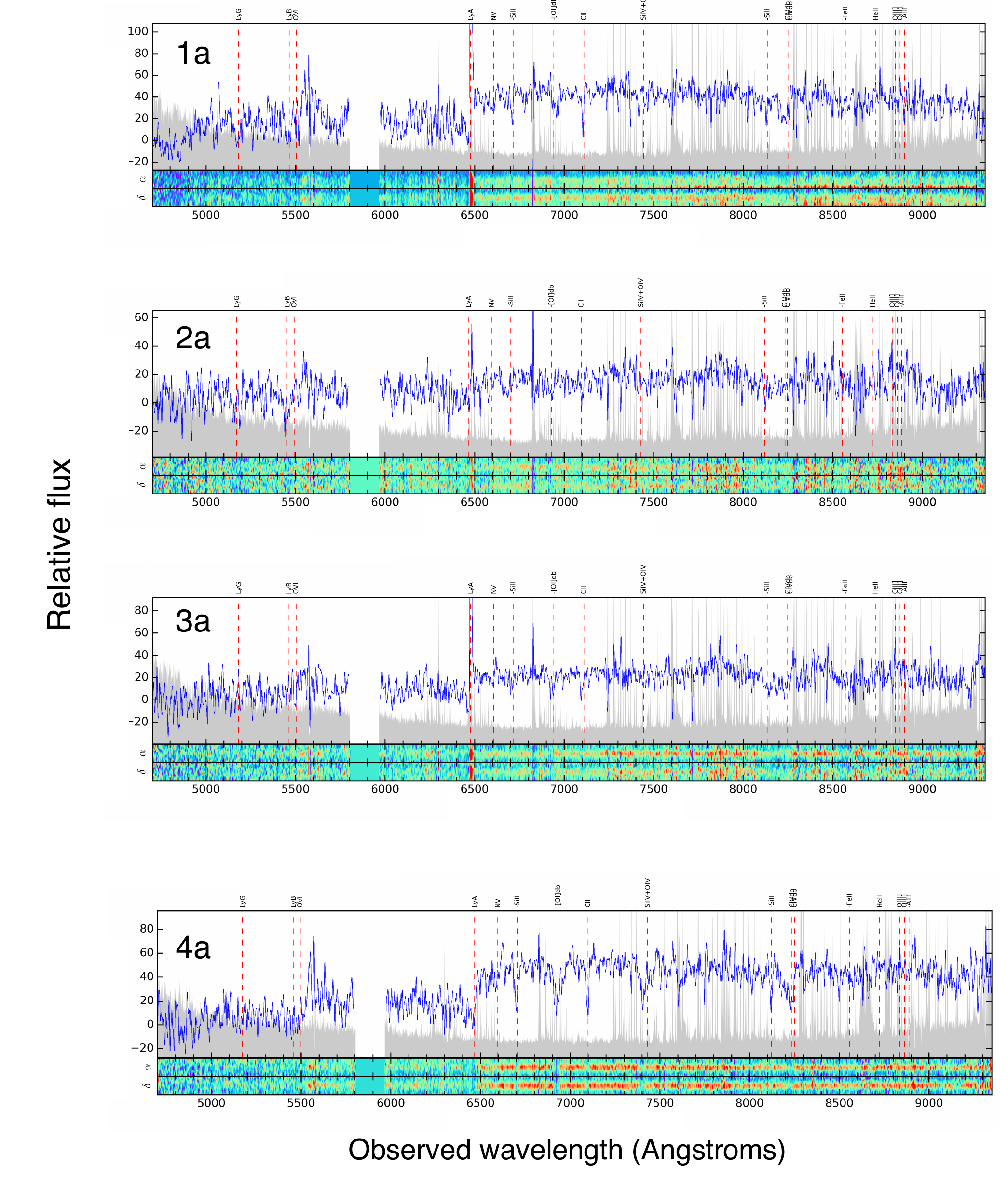}
\caption{VLT/MUSE 1D spectra of the four main galaxy members of the group associated with the ALMA source. For each source, we show the spectrum of highest S/N ratio among those corresponding to the different lensing images. Sources ID \#1 and \#3 clearly have Lyman-$\alpha$ in emission.  Source ID \#4, which is the closest to the ALMA source, has Lyman-$\alpha$ in absorption and displays a very prominent Lyman break in its spectrum. The shape of the Lyman-$\alpha$ line is less clear for source ID \#2, as it is contaminated by a sky line. The pseudo-2d spectra at the bottom of each panel are extracted in square apertures with a 3.2~arcsec side (i.e., 16 MUSE pixels) and projected in the RA (alpha) and Dec (delta) directions as indicated in the y-axes of the figures.} \label{fig:spec}
}
\end{figure*}

Two galaxies in the main group (ID \#1 and \#3) have Lyman-$\alpha$ in emission, with a clear P-Cygni profile,  according to the MUSE spectra (Fig.~\ref{fig:spec}). The signal-to-noise (S/N) ratio of the spectra is high enough to enable the detection of rest UV absorption lines, such as CII~$\lambda 1335$, SiII~$\lambda 1526$ and CIV~$\lambda\lambda 1548,1551$.

Source ID \#4, instead,  shows Lyman-$\alpha$ only in absorption (Fig.~\ref{fig:spec}), with a very prominent Lyman break. Several other UV rest absorption lines are also clearly identified. For source ID \#2, the Lyman-$\alpha$ line is contaminated by a sky line, so it is unclear whether it has an emission component. Nonetheless, the continuum break, as well as some UV absorption lines (CII, SII), are evident and allowed us to safely determine the galaxy spectroscopic redshift.

The spectroscopic redshifts of the different MUSE galaxies  (ID \#1 to \#4) allow us to estimate the rest-frame group velocity dispersion. We obtained $\sigma_r\approx 300 \pm 160 \, \rm km/s$, which indicates that this is likely an unrelaxed galaxy group (although, strictly speaking, the term `unrelaxed' refers to non-Gaussian radial velocity distributions, which cannot be determined for our galaxy group with only four members).

\begin{figure*}[h!]
\center{
\includegraphics[width=1.0\linewidth, keepaspectratio]{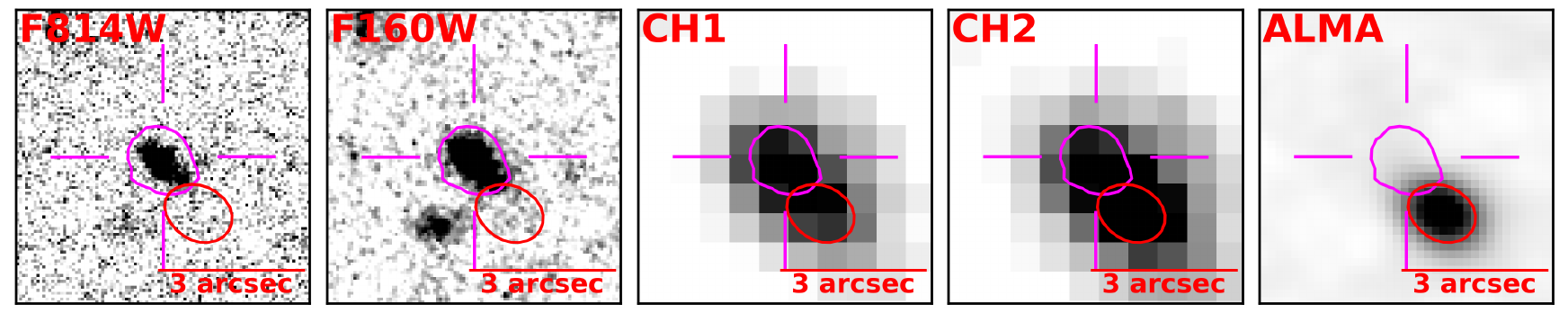}
\caption{Postage stamps of MUSE source ID \#4a and its associated ALMA detection. In all frames, the magenta and red contours show the MUSE and ALMA continuum emission regions, respectively.} \label{fig:almasrc}
}
\end{figure*}

Independently, we found an ALMA Band 6 source which is associated with the MUSE source ID~\#4 (Fig.~\ref{fig:almasrc}). The total measured ALMA 1.2~mm flux densities are $S_\nu(1.2 \, mm) = 9.50\pm 0.12, 3.83 \pm 0.11$ and $3.29 \pm 0.12$~mJy, in the components a, b and c, respectively, making it a very secure detection in all cases.   The ALMA source has been detected with the \textit{Spitzer} Infrared Array Camera \citep[IRAC;][]{faz04}, but is detected neither with MUSE nor \textit{HST}  (it has a colour F160W-[4.5]$>4$).  

\begin{figure}[ht!]
\center{
\includegraphics[width=1\linewidth, keepaspectratio]{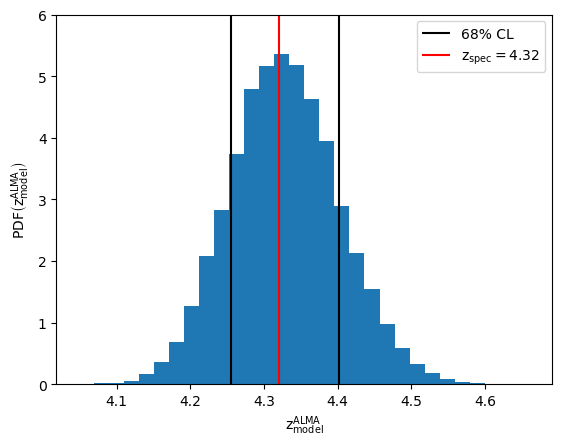}
\caption{Probability density distribution of the ALMA source redshift based on lensing analysis \citep{cam21}. This redshift distribution indicates that it is very likely that the ALMA source belongs to the same galaxy group as the MUSE galaxies at $z=4.32$}. \label{fig:almazlens}
}
\end{figure}

The ALMA source redshift can be obtained through the lensing model analysis \citep{cam21}. Its lensing-derived redshift is $z_{lens}=4.32^{+0.08}_{-0.06}$ (Fig.~\ref{fig:almazlens}), which indicates that the ALMA source is very likely part of the same galaxy group as the MUSE sources. In particular, the angular separation between the ALMA source and MUSE ID \#4 implies that these sources are at a projected distance of $2.07 \pm 0.88$~kpc in the source plane, which suggests that they might be two regions of the same galaxy. Indeed,  other authors have reported intrinsic offsets of about 0.4~arcsec between the rest-frame UV and FIR emitting regions of a same galaxy \citep[e.g.,][]{che15}, and these offsets can lead to separations of more than 1~arcsec in the image plane \citep{fuj16}. Nevertheless, we will analyse these two sources separately throughout this paper and discuss further the implications of being the same galaxy in \S\ref{sec:disc}.

\begin{deluxetable*}{lclllcccr}[b!]
\tablecaption{List of MUSE sources belonging to the galaxy group at $z\approx4.32$ in their three lensed components. The coordinates and properties of the associated ALMA source are also listed in each case.  \label{tab:sumtab}} 
\tablecolumns{9}
\tablenum{1}
\tablewidth{0pt}
\tablehead{ 
\colhead{ID} &
\colhead{$z_{spec}$} & 
\colhead{RA} & 
\colhead{DEC} & 
\colhead{Ly$\alpha$} & 
\colhead{$M_{\ast} \, \rm [\times 10^{9} \,  M_{\odot}]$} &
\colhead{$\rm SFR \, [M_{\odot} \, yr^{-1}]$} &
\colhead{$\rm sSFR \, [\times 10^{-9} \,yr^{-1}]$} &
\colhead{$\mu \pm 68\% \, \rm CL$}
}
\startdata 
1a   & 4.3278 & $15.7324575$ & $-49.2500982$ & E & --- & --- & ---  & $10.50^{+2.53}_{-1.62}$ \\
2a   & 4.3175 & $15.7332889$ & $-49.2514950$ & A?\tablenotemark{b} & $0.02\pm0.01$ & $1.06\pm0.24$ & $53.0\pm28.6$  & $5.64^{+0.36}_{-0.34}$  \\ 
3a   & 4.3275 & $15.7343748$ & $-49.2519405$ & E & $3.18\pm1.48$ & $3.70\pm0.79$ & $1.17\pm0.59$  & $5.19^{+0.30}_{-0.29}$\\ 
4a   & 4.3196 & $15.7323007$ & $-49.2522379$ & A & $40.1^{+65.3}_{-36.7}$ & $23.48\pm2.14$ & $0.59\pm0.75$ & $8.25^{+0.63}_{-0.59}$\\ 
ALMA-a & ---   & $15.7320058$ & $-49.2525287$ & --   & $77.7^{+390.1}_{-49.4}$ & $221\pm93$ & $2.84^{+8.26}_{-2.57}$ &  $9.77^{+0.79}_{-0.73}$\\ 
5\tablenotemark{a}     & 4.3181 & $15.7580432$ & $-49.2739080$ & E & --- & --- & --- & $4.80^{+0.18}_{-0.21}$ \\ 
6\tablenotemark{a}     & 4.3187 & $15.7623243$ & $-49.2800099$ & E & --- & --- & --- & $23.99^{+53.56}_{-14.67}$\\ 
\hline 
1b   & 4.3273 & $15.7262164$ & $-49.2534759$ &  E &  --- & --- & --- & $4.19^{+0.26}_{-0.25}$ \\
2b   & ---    & $15.7275328$ & $-49.2545678$ & A?\tablenotemark{b} &  --- & --- & --- & $5.60^{+0.36}_{-0.33}$ \\
3b   & 4.3269 & $15.7281160$ & $-49.2554323$ & E  & $1.76\pm0.49$ & $2.20\pm0.48$ & $1.25\pm0.43$ & $6.15^{+0.44}_{-0.39}$\\
4b   & ---    & $15.7285033$ & $-49.2541646$ & A & --- & --- & --- & $2.85^{+0.42}_{-0.36}$\\
ALMA-b & ---   & $15.7288345$ & $-49.2540943$ & &   --- & $296\pm124$ & --- & $2.53^{+0.45}_{-0.39}$ \\ 
\hline 
1c   & 4.3273 & $15.7123036$ & $-49.2593173$ & -- &  $0.09\pm0.08$ & $4.13\pm 0.86$  & $45.9\pm41.9$ & $3.98^{+0.15}_{-0.15}$ \\
2c   & ---    & $15.7134793$ & $-49.2602988$ & -- &  $0.009\pm0.004$ & $0.45\pm0.15$ & $50.0\pm27.8$ &  $4.18^{+0.16}_{-0.16}$\\
3c   & 4.3289 & $15.7147336$ & $-49.2606804$ & -- & $1.10\pm0.56$ & $1.62\pm0.34$  & $1.47\pm0.87$ &  $4.36^{+0.18}_{-0.18}$\\
4c   & ---    & $15.7128864$ & $-49.2607064$ & -- & $0.92\pm0.23$ & $4.40\pm0.94$ & $4.78\pm1.53$ &  $4.13^{+0.17}_{-0.17}$\\
ALMA-c & ---   & $15.7128881$ & $-49.2608024$ & -- & $58.1^{+288.4}_{-30.7}$ & $201\pm84$ & $3.46^{+6.94}_{-3.12}$&   $4.18^{+0.18}_{-0.17}$\\ 
\enddata
\tablenotetext{a}{Source at very similar redshift to the others, but farther away, so their association with the main MUSE galaxy group (sources ID \#1 to 4) is unclear.}
\tablenotetext{b}{There is a sky line on top of the Lyman-$\alpha$ line in the MUSE spectrum of source 2, so it is unclear whether this line is only in absorption or has a component in emission.}
\tablecomments{Column 5 indicates whether Ly$\alpha$ is in emission (E) or only in absorption (A). The SFRs in column 7 have been obtained from the extinction-corrected, rest-UV flux of each galaxy, except for the ALMA source, for which the dust-obscured SFR has been derived with MAGPPHYS. All stellar masses (column 6) and SFRs (column 7) are corrected for lensing magnification \citep[using column 9;][]{cam21}. }
\end{deluxetable*}

\subsection{Properties of the group galaxies\label{subsec:prop}}

\subsubsection{The lensed MUSE sources}

\begin{figure*}[ht!]
\center{
\includegraphics[width=1.0\linewidth, keepaspectratio]{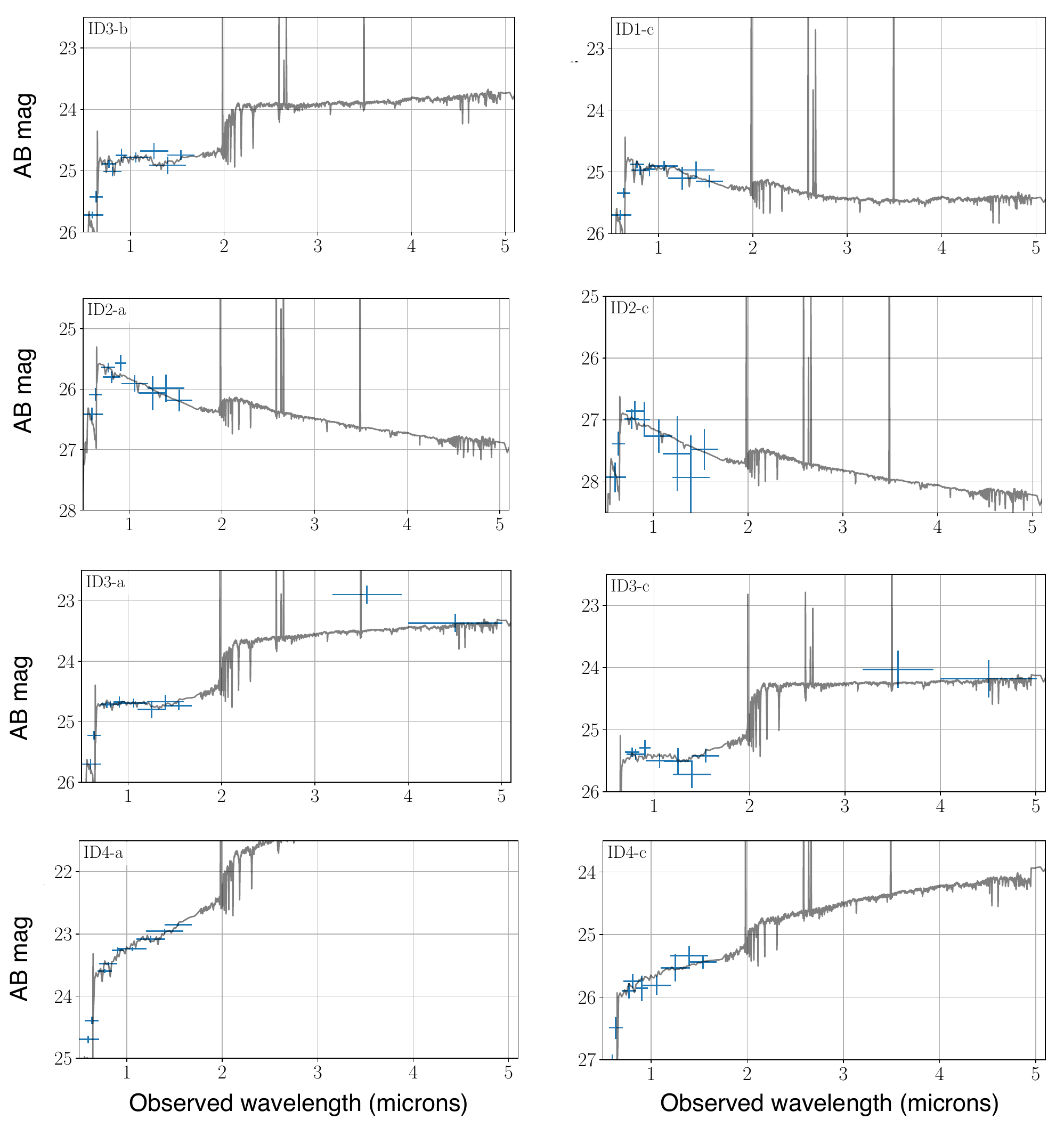}
\caption{Best-fit SEDs for the four MUSE sources in the galaxy group at $z=4.32$, in all the lensing components in which we have performed the SED fitting (those with photometry which is not significantly contaminated by the foreground cluster).} \label{fig:seds}
}
\end{figure*}

We performed the optical/near-IR spectral energy distribution (SED) fitting of all the MUSE sources of the main group (sources ID \#1 to \#4) at their fixed spectroscopic redshifts, in order to recover their  main properties (Fig.~\ref{fig:seds}). For this, we made use of the SED fitting code \textsc{LePhare} \citep{arn99,ilb06}. We considered \citet{bc03} templates corresponding to a single stellar population and different exponentially declining star formation histories (SFHs), both with solar ($\rm Z_\odot$) and sub-solar ($0.2 \, \rm Z_\odot$) metallicities. To account for dust extinction we convolved the templates with the \citet{cal00} reddening law for different extinction values from $E(B-V)=0$ to $1$.

We applied the SED fitting over 11 \textit{HST} broad bands for sources in the components a and c, except for source ID \#1a because its photometry is contaminated by a foreground-cluster member. The system component b is also close to cluster member galaxies, and thus the photometry is substantially contaminated for most sources (except for object ID \#3b). We only incorporated IRAC photometry for source \#3 (a,c), as only for this object is the IRAC photometry not significantly blended. This source shows a clear IRAC flux excess at $3.6 \, \rm \mu m$, indicative of prominent H$\alpha$ emission at its redshift \citep[e.g., ][]{cap17}.  Taking this into account, we checked the SED fitting with both IRAC bands and only $4.5 \, \rm \mu m$ in addition to the \textit{HST} bands, and found that the results were consistent (we ran \textsc{LePhare} with emission lines, so the code can ``recognise'' the presence of H$\alpha$).

The SED best fitting results indicate that source ID \#1 has a preference for a solar metallicity template, while sources ID \#2 and \#3 have best fit solutions with sub-solar metallicities. The output is less clear for source ID \#4, which has best solution with solar metallicity in component a, while prefers a $0.2 \, \rm Z_\odot$ in component c. This might indicate that the true metallicity of this source is intermediate between $0.2 \, \rm Z_\odot$  and $\rm Z_\odot$. All sources except source ID \#4 have very low or no internal extinction. Even for source ID \#4 the extinction is modest, with $A_V < 1$~mag. This non-negligible extinction can explain why this source is the only one in the group with confirmed Lyman-$\alpha$ in absorption and the most prominent Lyman break among our group sources.

\textsc{LePhare}'s output also provides us with stellar mass estimates (Table~\ref{tab:sumtab}). The values derived for sources ID \#2 and \#3 are consistent within the error bars for the three different lensed components. Instead, the stellar mass of source ID \#4 appears more loosely constrained, with a value between $\sim 10^9$ and  $\sim 4 \times 10^{10} \, \rm M_\odot$, but in any case it is one of the most massive galaxies of our group.

We also independently derived star formation rates for all our galaxies using  their extinction-corrected, rest-frame UV fluxes (based on the F105W filter photometry) and the corresponding \citet{ken98} formula. For sources ID \#1,2,3 the SFR have a value of at most a few $\rm M_\odot/yr$. As for the stellar mass, the SFR of source ID \#4 is more difficult to constrain, but we find that its value is between $\sim 4.4$ and $\sim 23.5 \, \rm M_\odot/yr$. 

For source ID \#3, we can independently derive an SFR from the H$\alpha$ excess in the IRAC 3.6$\rm \mu m$ band. Following the methodology described in \citet{cap17}, we obtained $(28.2 \pm 7.9)$ and $(4.74\pm1.33) \, \rm M_\odot/yr$ for sources ID \#3a and c, respectively (both values are corrected for magnification). These SFRs are significantly larger than those obtained from the rest UV fluxes (Table~\ref{tab:sumtab}). These differences suggest that the extinction correction derived from the SED fitting of this source is likely underestimated (a minor difference in the $E(B-V)$ value would produce a significant change in the inferred intrinsic UV flux). Nevertheless, these higher SFRs derived from the H$\alpha$ line do not qualitatively change the classification of this source on the SFR versus $M^\ast$ plane, as we discuss below.

All stellar masses and SFRs quoted in Table~\ref{tab:sumtab} have been corrected for lensing magnification. The error bars on the stellar masses and SFRs incorporate the lensing magnification errors in quadrature, but these are almost always a minor component of the total error budget in these quantities. 

In Table~\ref{tab:sumtab} we also quote the specific SFR (sSFR) for our galaxies. This parameter has the important advantage of being basically independent of lensing magnification (this is strictly true assuming that the magnification is constant throughout a galaxy). Therefore the sSFR determination should be considerd more robust than the SFR and stellar mass separately. 

\begin{figure*}[ht!]
\center{
\includegraphics[width=1.0\linewidth, keepaspectratio]{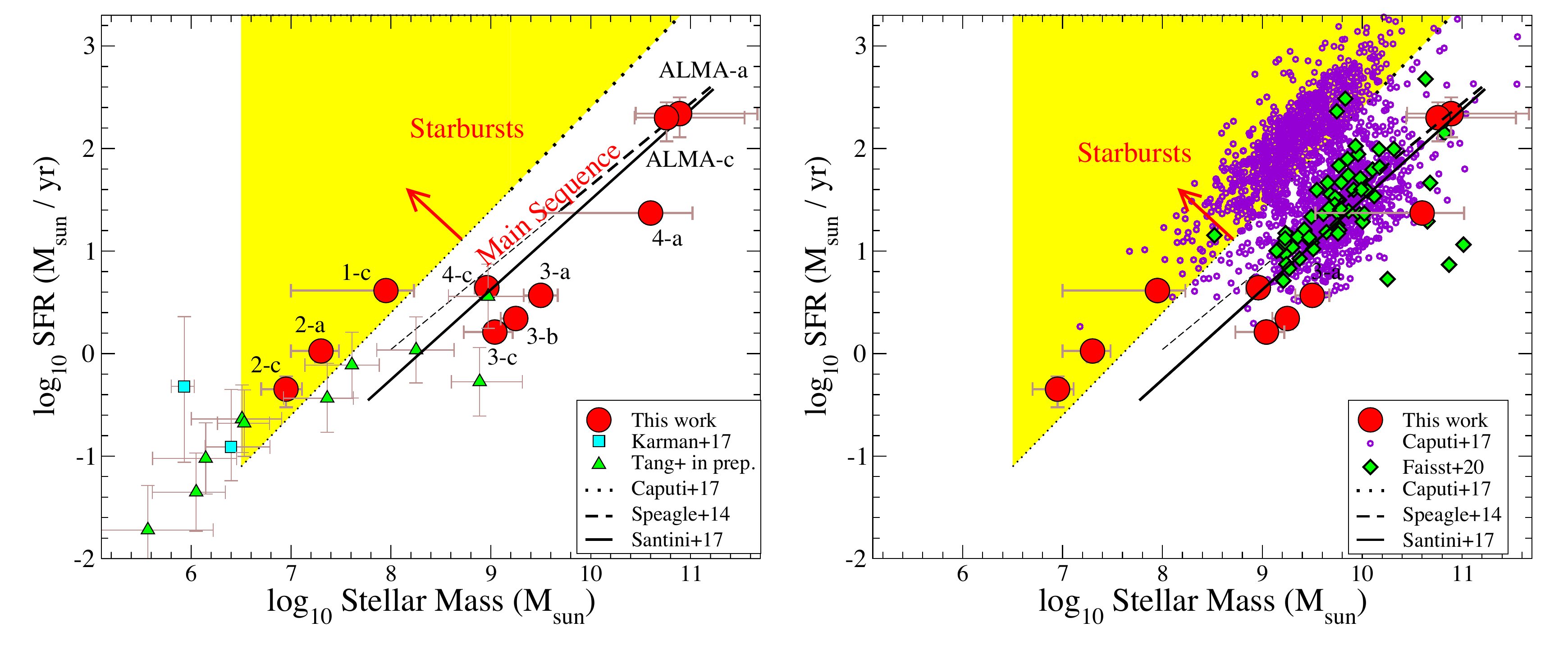}
\caption{Location of our group sources on the SFR vs. $M^\ast$ plane, in comparison to other sources from the literature at redshifts $4<z<5$.   {\em Left:} comparison to other lensed galaxies. {\em Right:} comparison to galaxies in the field, particularly prominent H$\alpha$ emitters \citep{cap17} and ALMA galaxies \citep{fai20}. In both panels, 
we indicate the average position of the star-formation main sequence at the same redshifts, as obtained by \citet{spe14} (dashed line) and \citet{san17} (solid line; determined by the analysis of lensing fields).  The dotted line shows the lower envelope for the starburst classification at $z\sim 4.3$ \citep{cap17}. In all cases, thick lines are used for the stellar mass ranges in which these relations are known to be valid, while thin lines indicate linear extrapolations at lower stellar masses.  For clarity, the source labels have been removed in the right panel.} \label{fig:sfrstm}
}
\end{figure*}

Figure~\ref{fig:sfrstm} shows our MUSE group sources on the SFR versus $M^\ast$ plane, along with other data points corresponding to lensed galaxies in other cluster fields and field galaxies, all at $4<z<5$. For reference, we also show the location of the star formation main sequence as well as starburst galaxies, as derived in previous works, albeit only for galaxies with stellar masses $\gsim 10^8-10^9 \, \rm M_\odot$ \citep{spe14,cap17,san17}.  Note that the starburst line in Fig.~\ref{fig:sfrstm} is the lower envelope of the starburst locus, defined by $\rm log_{10} (sSFR(yr^{-1})) \geq -7.60$  (see \citet{cap17}). 

From the location of our galaxies in this diagram we see that sources ID \#3 and \#4 can be classified as main-sequence galaxies. Interestingly, source ID \#3 would still be classified as a main-sequence galaxy considering the SFRs inferred from the H$\alpha$ excess in the IRAC 3.6$\rm \mu m$ band.  Sources ID \#1 and \#2, instead,  lie above the starburst division line. Note, however, that we should take this classification with care because the main-sequence/starburst division is not known at stellar masses $\lsim 10^8 \, \rm M_\odot$ and a linear extrapolation from higher stellar masses might not be strictly valid.  In any case, the sSFR derived for sources ID \#1 and \#2 imply that the doubling time for their stellar masses is only $~2\times 10^7$~years, which is consistent with the short timescales expected for star-formation episodes in local starburst galaxies \citep[e.g.,][]{kna09}.

None of the MUSE sources at $z=4.32$ is individually detected by ALMA, except for the association between the ALMA source discussed here and source MUSE ID \#4. An ALMA stacking analysis of the MUSE group sources ID \# 1, 2  and 3 in the three lensed components does not result in a signal either, as probably expected, indicating that the average total infrared luminosity of these sources must be below a few $\times 10^{10} \, \rm L_\odot$ (after correcting for lensing magnification).

\subsubsection{The lensed ALMA source}

We also estimated the dust-obscured SFR for the ALMA source, using two different methods. In the first approach, we corrected the ALMA 1.2~mm flux densities for the lensing magnification in each component (see Table~\ref{tab:sumtab}) and then we scaled them taking into account empirical sub-millimetre galaxy templates, following the same methodolgy as in \citet{cap14}. We obtained total infrared luminosities ranging between $\rm L_{\rm TIR}=(1.23 \pm 0.52)\times 10^{12} \, \rm L_\odot$ (component c)  and  $(2.32 \pm 0.95)\times 10^{12} \, \rm L_\odot$  (component b). Therefore, the ALMA source can be classified as an ordinary ultra-luminous infrared galaxy (ULIRG), with a luminosity more typical of the ULIRGs known at $z=2-3$ than the luminosities of the higher-redshift sub-/millimetre sources commonly studied in the pre-ALMA era.  The derived SFRs using the \citet{ken98} formula corresponding to infrared luminosities are $\mathrm SFR\approx 130-250 \, \rm M_\odot/yr \, (\pm 42\%)$ (after correction to a Chabrier IMF).

As a second, independent analysis, we considered the ALMA 1.2~mm flux densities, along with \textit{Herschel}/SPIRE flux densities at 250, 350 and 500$\rm \mu m$ to perform the SED fitting of the ALMA source dust emission using the high-$z$ extension of the code \textsc{MAGPHYS} \citep[][]{dac08}. The \textit{Herschel} detections are most secure for the ALMA source in component a (ALMA-a), for which the three SPIRE bands yield a $>3 \sigma$ detection. ALMA-b was detected at $3\sigma$ at 500$\rm \mu m$, while ALMA-c was only detected above $2\sigma$ at 350$\rm \mu m$. In the case of non-detections, an upper limit of $\sim 15$~mJy ($3\sigma$) can be placed for their peak far-IR flux density in the SPIRE bands.

The output of the \textsc{MAGPHYS} SED fitting is consistent with our first total infrared luminosity estimates for the ALMA source. After correcting for lensing magnification, the values derived with \textsc{MAGPHYS} range between $\rm L_{TIR}=(1.86\pm 0.89)\times 10^{12} \, \rm L_\odot$ (component c)  and  $(2.73\pm 1.07)\times 10^{12} \, \rm L_\odot$  (component b). These values correspond to dust-obscured SFRs ranging between $\approx 200$ and $300 \, \rm M_\odot/yr \, (\pm 42\%)$ (after correction to a Chabrier IMF). These values are in reasonably good agreement with those obtained from our first estimation method for the total infrared luminosities.

We also estimated dust temperatures for these sources, following the methodology of \citet{dud20}. We obtained  $T_{\rm dust} \approx 30.1-34.8 \, \rm K$, which is consistent with the median temperature derived by \citet{dud20} for sub-millimetre galaxies with $\rm L_{TIR}\sim 3.4 \times 10^{12} \, \rm L_\odot$ at $z\sim 2.5$.

\begin{figure}[ht!]
\center{
\includegraphics[width=1.0\linewidth, keepaspectratio]{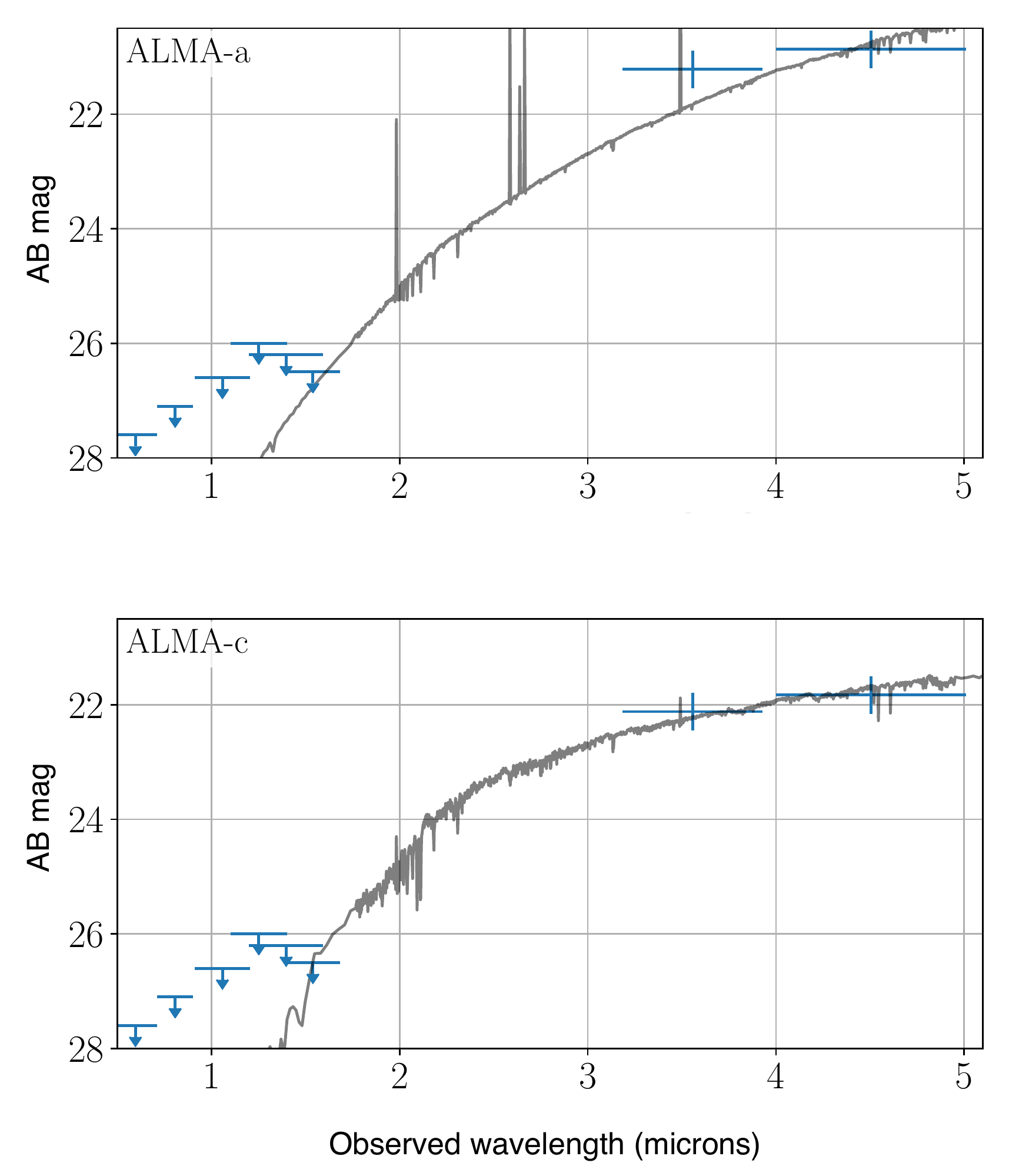}
\caption{Best-fit SEDs for the ALMA source at $z=4.32$,  in  the two lensing components in which we have performed the SED fitting.} \label{fig:almastsed}
}
\end{figure}

At wavelengths tracing the stellar emission, the ALMA source is only detected in the IRAC 3.6 and 4.5~$\rm \mu m$ bands. Based on two secure data points, and  {\textit{HST}} flux upper limits, it would normally be difficult to do a proper SED fitting. However, in our case, we exploit the fact that we have an independent redshift determination for the ALMA source. We fit its SED running \textsc{LePhare} at the known fixed redshift to obtain an estimate of its stellar mass for two  lensed components (see Fig.~\ref{fig:almastsed} and Table~\ref{tab:sumtab}). We obtained that the lensing corrected stellar mass of the ALMA source is between $\sim5.8$ and $7.8 \times 10^{10} \, \rm M_\odot$. The stellar mass values obtained for the two lensed components are in a remarkably good agreement taking into that the SED fitting is mostly constrained by flux upper limits and that the IRAC photometry is somewhat blended with that of source ID \#4.

With the SFR and stellar mass in hand, we can locate the ALMA galaxy on the SFR versus $M^\ast$ plane (Fig.~\ref{fig:sfrstm}). As it is clear from this plot, the ALMA source is a main-sequence galaxy. Therefore, if MUSE source ID \#4 and the ALMA source were two regions of the same galaxy, then the total system would still constitute a main-sequence galaxy.

\section{Discussion} \label{sec:disc}

The existence of deep MUSE data on a lensing field reveals that the rather typical ALMA galaxy discussed in this paper is not alone. There are at least four star-forming galaxies associated with the ALMA source at $z=4.32$, as confirmed by the three lensing components. 

Previous works have shown that sub-/millimetre sources can trace active environments with multiple star-forming galaxies \citep[e.g.,][]{ume15,mil18,ote18,fry19} and even be tracers of cosmic web filaments \citep[e.g.,][]{ume19}. In most of these studies the system is dominated by an intrinsically bright sub-/millimetre galaxy.  Gravitational lensing allows us to show that this property of tracing active environments is extended to less luminous systems. \citet{kne04} reported a lensed galaxy group which includes an intrinsically sub-mJy sub-millimetre galaxy at $z\sim 2.5$. Here we report the discovery of another galaxy group of a similar kind, albeit much more compact and at a significantly higher redshift, $z=4.32$.

The MUSE spectra, combined with the lensing effect, has been fundamental to this discovery. They provide the spectroscopic redshift confirmation for the group galaxies, as well as the necessary constraints for the lensing model to derive a secure redshift for the ALMA source and confirm its association with the MUSE galaxy group.
 
In spite of the dusty SFR derived for the ALMA source, which is much higher than the SFR of any other galaxy in the group, it is classified as a main-sequence galaxy on the SFR-$M^\ast$ plane. This should not be surprising, as other authors found that many high-$z$ sub-millimetre galaxies lie on the star-formation main sequence \citep[e.g.,][]{mic12,sch16,lee17,mic17,sch18}. Source MUSE ID \#4 is also classified as a main-sequence galaxy. If both sources were in fact regions of a single galaxy, the combination of both would still result in a main-sequence galaxy.

Although the derived SFRs of the MUSE sources are low in comparison with that of the ALMA source  (in total they make for $\sim$10\% of the ALMA galaxy SFR),  in some cases these SFRs are still relatively important taking into account the low stellar masses.  Indeed, there are two starburst galaxies (ID \#1 and \#2) within the MUSE galaxy group. These two galaxies are expected to double their stellar masses in only $~2\times 10^7$~years, as derived from their sSFRs. Instead, MUSE source ID \#3, like source ID\#4, lies on the star-forming galaxy main sequence.  

 The classification of galaxies in the SFR versus $M^\ast$ plane is poorly known for stellar masses $\lsim 10^9 \, \rm M_\odot$, as this region of parameter space is currently only constrained by a small amount of galaxies studied in lensed fields. We do not know whether the main-sequence/starburst classification, as it is usually defined at $\gsim 10^9 \, \rm M_\odot$, is still valid at lower stellar masses. Nevertheless, we note that the statement that our MUSE sources ID \#1 and \#2 are starbursts is mainly based on their stellar-mass doubling times, which is of only $~2\times 10^7$~years. These short timescales are similar to those characterising star-formation episodes in local starburst galaxies \citep[][]{mih94,kna09}.

The presence of two starburst galaxies in this galaxy group with only 4-5 members makes for a starburst percentage of 40-50\%. Although this is based on very low statistics, this percentage is much higher than the overall starburst percentage known among star-forming galaxies at any redshift. This is perhaps not surprising taking into account the nature of our galaxy group. This is a compact galaxy group with a high velocity dispersion, in which the object interaction is likely triggering new star-formation episodes. The effect is most drastic in the lowest mass galaxies, which is in line with the finding that the starburst fraction increases with decreasing stellar mass at different redshifts \citep{bis18}.  It remains to be seen how ubiquitous these systems are, and if other similar systems also display such an enhanced fraction of starburst galaxies. Collecting statistically larger samples of these compact groups at different redshifts is necessary to draw a more general conclusion on this issue.

\acknowledgments

Based in part on observations carried out with ESO Telescopes at the Paranal Observatory under ESO programme ID 0102.A-0266. Also based on observations carried out by NASA/ESA {\em Hubble Space Telescope}, obtained and archived at the Space Telescope Science Institute. This paper makes use of the following ALMA data: ADS/JAO.ALMA\#2018.1.00035.L. ALMA is a partnership of ESO (representing its member states), NSF (USA) and NINS (Japan), together with NRC (Canada), MOST and ASIAA (Taiwan), and KASI (Republic of Korea), in cooperation with the Republic of Chile. The Joint ALMA Observatory is operated by ESO, AUI/NRAO and NAOJ.

We thank an anonymous referee for a constructive report. We thank Ian Smail for discussion on the paper.  KIC, GBC and SD  acknowledge funding from the European Research Council through the award of the Consolidator Grant ID 681627-BUILDUP. KK acknowledges the support by JSPS KAKENHI Grant Number JP17H06130  and the NAOJ ALMA Scientific Research Grant Number 2017-06B. YA acknowledges support by NSFC grant 11933011. KKK acknowledges support from the Swedish Research Council and the Knut and Alice Wallenberg Foundation. CG acknowledges support through grant no.~10123 of the Villum Fonden Young Investigator Programme. GEM acknowledges the Villum Fonden research grant 13160 ``Gas to stars, stars to dust: tracing star formation across cosmic time'' and the Cosmic Dawn Center of Excellence funded by the Danish National Research Foundation under the grant No. 140. HU acknowledges support from JSPS KAKENHI grant (20H01953). EV acknowledges the financial support provided by the INAF  for ``interventi aggiuntivi a sostegno della ricerca di main-stream''  Main-Stream (1.05.01.86.31).

%

\vspace{5mm}
\facilities{ALMA, VLT, \textit{HST}, \textit{Spitzer}, \textit{Herschel}}


\software{SExtractor, IRAF, LePhare, MAGPHYS}





\end{document}